\documentclass[12pt]{iopart}
\usepackage{graphicx, subfigure,hyperref,mathrsfs}
\usepackage{color}
\usepackage[margin=10pt,font=small,labelfont=bf]{caption}
\def\clr {\color{black}}
\def\clrr {\color{black}}
\def\blam{\bar \lambda}

\def\text{}
 \newcommand{\beq}{\begin{equation}}
 \newcommand{\eeq}{\end{equation}}
\begin{document}
\title{Conserved mass models with stickiness and chipping}

\author{Sourish Bondyopadhyay and P. K. Mohanty}

\address{Theoretical Condensed Matter Physics Division,\\ Saha Institute of
 Nuclear Physics,1/AF Bidhan Nagar, Kolkata-700064, INDIA.}
\ead{sourish.bondyopadhyay@saha.ac.in and pk.mohanty@saha.ac.in}
\date{\today}
\vskip 2.cm
\begin{abstract}
We  study a chipping model in one dimensional periodic lattice with
continuous mass, where a  
fixed  fraction of the mass  is chipped off from a site and  
distributed randomly among the departure site and its neighbours; 
 the remaining mass sticks to the site. In the asymmetric  version,
the  chipped off mass is distributed among  the site and the   
right neighbour, whereas  in the  symmetric version  the   
redistribution occurs among  the two neighbours. The steady state
mass distribution of the  model  is obtained  using a perturbation method 
for both  parallel and random sequential  updates.
In  most cases, this perturbation theory provides a steady state  
distribution with reasonable accuracy. 
\end{abstract}
\vspace{2pc}

\maketitle
\section{ Introduction :}~\indent 
Most systems in nature are in non-equilibrium states \cite{noneqbook}, in  a
way 
that the  accompanying fluxes of mass, energy,  or spin {\it etc.} are irreversible. 
Unlike their  equilibrium counterparts where the stationary state is
characterized 
by the Gibbs measure,  these systems usually  reach  different and novel
stationary 
states depending on  the dynamics of the microscopic constituents. Several 
non-equilibrium lattice  models have been  proposed recently
\cite{noneqmodels} to investigate  
the unusual steady state  distributions,  spatio-temporal correlations  and 
possibility of macroscopic  collective phenomena.

One of the simple non-equilibrium models is  the mass transport model  where 
each site of a lattice is associated with  discrete mass (particles) 
following a dynamics that involves  aggregation, fragmentation, adsorption 
or desorption \cite{AggDiffDisl}. 
Like  the zero range process \cite{ZRP},  
interestingly,  many of these  model systems  undergo a  condensation phase transition as  
the mass density of the system is  increased.  Study of these models  have
generated considerable interest among physicists, as  a wide variety of 
systems exhibit basic microscopic mechanism 
similar to that of the simple mass transport models. These include  
colloidal suspensions \cite{colloid}, polymer gels \cite{polymer1},
river networks \cite{river}, traffic models
\cite{traffic},  and wealth distribution 
\cite{wealth}  {\it etc}.

A continuous version of the mass 
transport model is proposed recently \cite{Rajesh} and some of its 
variations have also been studied \cite{condtransport,FSS,FSS2}. 
Many of the mass transport models are known to evolve into a non-equilibrium  
steady state that has  product measure.  {\clrr A generic  criterion for 
factorized steady state has been derived \cite{FSS,FSS2} for mass
transport models.}
In our effort to study models where  the steady state is not 
factorized and only a little is known analytically,  we develop a
perturbation approach  that provides an approximate form of the 
steady state distribution.

In this article we introduce  {\it stickiness}, quantified by a parameter
$\lambda,$   
to the continuous mass transport 
models. At each site,  $(1-\lambda)$ fraction of the mass is chipped off 
(thus $\lambda$ fraction of the mass  sticks to the site), which is then 
redistributed either asymmetrically, {\it i.e.} among the departure site and its
 right neighbour, 
or symmetrically, {\it i.e.} among  the  two neighbours. Accordingly, the 
model is referred to  as  asymmetric or symmetric   sticky chipping model 
respectively.   We use a novel perturbation approach to calculate the steady
state 
mass distribution of these models, for both  parallel and 
random sequential updates.  Although in this  perturbation approach, we 
have ignored two and three point spatial correlations, the 
mass distributions calculated up to $2^{nd}$ order are 
strikingly close to the same  obtained from Monte Carlo 
simulations. 

The article is organized as follows. The model and the perturbation method  are
outlined in section 
\ref{model}. In section \ref{ASCM} we  study the asymmetric version of the model
and obtain the steady state distribution. The symmetric version of the
model, where the chipped off mass is distributed among both the neighbours, 
is discussed in section \ref{SSCM}.  Finally we discuss 
the main results  in  section \ref{summary}. The  perturbation  results 
 up to second order for all these models are listed  in the Appendix.

\section {The Model :} ~
\label{model}
\indent  The model is defined on a one dimensional periodic lattice with sites
labeled by $i=1,2, \dots ,L$.  A continuous  mass $x_i$,  
associated at each site $i,$ evolves according to the following dynamics. At each site $i$, 
$(1-\lambda)$ fraction of the mass is chipped off (thus, $\lambda$ 
fraction of the mass sticks to the site) and then it is distributed 
among the departure site $i$ and its neighbours 
$(i\pm 1).$  In this article we study  two different  versions; 
in the asymmetric  sticky chipping model  (ASCM)   the chipped off mass
$(1-\lambda)x_i$ is  distributed  randomly
among the  sites  $i$ and $(i+1),$ whereas in the 
symmetric   sticky chipping model (SSCM) it is distributed among 
the neighbours  $(i+1)$ and  $(i-1)$. 
Both  these  versions  are studied  using  parallel 
and  random sequential update rules.

 We must mention that   the steady state mass distribution of 
ASCM   with $\lambda =0$  has been obtained  
earlier by Rajesh 
$et. al.$ \cite{Rajesh} assuming that the steady state is
factorized. {\clr  This  product measure assumption  
turns out to be exact in case of parallel update and  an  
excellent  approximation (though not exact) in the random 
sequential case.  We will show  that  in presence of 
stickiness $(\lambda \ne0),$   neither ASCM nor SSCM can have 
a factorized steady state.  To construct  the 
steady state mass distribution $P(x)$, we use  a novel  perturbation 
approach. Although   the  spatial correlations are ignored here, the 
steady state distributions, calculated up to $2^{nd}$ order in $\lambda,$ are found to be in excellent agreement 
with the same obtained  from the  Monte Carlo simulations.} 
The  general principle of this approach 
is described in the following subsections.

\subsection{{\bf Perturbation approach I :}}~
\indent We try to construct the steady  state mass distribution $P(x)$ 
perturbatively, by expressing $P(x)$  as a power series in $\lambda, $ about
$\lambda=0$ ,
\begin{equation}
P(x)=P_0(x)+\lambda P_1(x)+\lambda^2 P_2(x)+\ldots=\sum_{k=0}^\infty \lambda^k
P_k(x) 
\label{Pxs}
\end{equation}
where the functions  $P_k(x)$  do not depend on $\lambda$.  
Here we have omitted the argument  $\lambda$  in $P(x)$  for notational convenience.
{\clr The perturbative expansion  can also be made  about any other 
$\lambda$  if $P_0(x)$  can be calculated  there.
   In the chipping models
discussed here, the 
total mass of the system and hence  the density $\langle x\rangle = \frac{1}{L}
\sum_i x_i$ is conserved. 
Without any loss of generality one can fix the average mass  to be unity. This imposes
a condition on $P(x)$ 
\begin{equation}
 \int_0^\infty dx ~ x P(x)=\langle x\rangle =1=\int_0^\infty dx~ P(x)~,
\label{eq:avrx=1}
\end{equation}
where the last equality stands for the normalization condition.
Now, for $\lambda=0$, $P(x)=P_0(x)$. Therefore, $P_0(x)$ satisfies {\it two}
conditions, 
\beq
\int_0^\infty dx\ P_0(x)=1 ~{\rm and} \int_0^ \infty dx\ x P_0(x)=1.
\label{C0}
\eeq 
Thus  for any other $k>0,$ using Eqs. (\ref{eq:avrx=1}) and (\ref{C0}) we have 
\begin{equation}
 \int_0^{\infty}dx\ P_k(x)=0 =\int_0^\infty dx \ x P_k(x) .
\label{Eq.pkb}
\end{equation}
The above constraints  can not be satisfied by any real positive function (as
$x\ge0$), and  thus one can not 
interpret  $P_k(x)$ as a probability density function.

The  perturbative corrections $P_k(x)$  can be obtained directly from 
knowing the moments of $P(x)$. First let us expand the moments {\clr
$\langle x^n\rangle =  \int_0^\infty dx\ x^n P(x)$  as a power series in $\lambda$, 
\begin{equation}
 \frac{\langle x^n\rangle}{(n+1)!} \equiv A^{(n)} =\sum_{k=0}^\infty \lambda^k A_k^{(n)}.
\label{Xs}
\end{equation}
Here $A_k^{(n)}$ are constant coefficients (independent of $\lambda$).   
The  usefulness of the factor $1/(n+1)!$,  used here for notational convenience, 
will be  clear as we  discuss specific problems. } 
$A_k^{(n)}$ can be determined from Eqs. (\ref{Pxs}) 
and (\ref{Xs}), 
\begin{equation}
A_k^{(n)}= \frac{1}{(n+1)!}\int_0^\infty dx \  x^n  P_k(x).
\label{eq:Ck}
\end{equation}
Since   $P_k(x)$ are constrained by Eqs. (\ref{C0})  and (\ref{Eq.pkb}) 
$A_k^{(n)}$ must satisfy, 
 \begin{equation}
 A_0^{(1)} = \frac{1}{2} ~~  {\rm and} ~~ A_k^{(1)} = 0~~\forall~~ k>0.
\label{Eq.bcA}
\end{equation}

 Once $A_k^{(n)}$  are known,  one can calculate  $P_k(x)$  as  
\begin{eqnarray}
P_k(x) =
\mathscr{L}^{-1} \left[
\sum_{n=0}^{\infty} (-s)^n (n+1)A_k^{(n)}\right],
\label{eq:Pks}
\end{eqnarray}
where  $\mathscr{L}^{-1}$  denotes the inverse Laplace transform.

\subsection{{\bf Perturbation approach  II :}}~
\label{ptII}
\indent Here we assume that the mass distribution $P(x)$ satisfies a
differential or an integral equation and  use the Laplace transform  
\beq Q(s)=\mathscr{L}\left[P(x)\right]=\int _0^ \infty dx \  e^{-sx}  P(x) \label{QP}\eeq
 which  usually  converts it into  
a differential or a transcendental equation in $Q(s).$ 
We proceed further by expanding $Q(s)$ as power series in $\lambda$ about
$\lambda=0$,
 \beq Q(s) =  \sum_{k=0}^\infty \lambda^k Q_k(s),\label{Qs}\eeq
and equate the coefficients of different powers of $\lambda$ order by order.
Finally one can find the distribution 
\beq P(x) = \mathscr{L}^{-1} \left[  Q(s) \right]=\mathscr{L}^{-1}
\left[\sum_{k=0}^\infty \lambda^k Q_k(s)\right].\eeq

These two  approaches are equivalent as it is evident from Eq. (\ref{eq:Pks}),
\begin{equation} 
 Q_k(s)= \mathscr{L} \left[ P_k(x)\right]=
\sum_{n=0}^{\infty}  (-s)^n (n+1)A_k^{(n)};
\label{Qk}
\end{equation}
for any particular problem we will use the approach whichever is convenient.

\section{Asymmetric Sticky Chipping Model (ASCM) :}~
\label{ASCM}
\indent In this section we study the asymmetric version of the model 
where $(1-\lambda)$ fraction of  mass $x_i$  at site $i$  is chipped 
off, only $r_i$ fraction  of  the chipped off mass is  then 
transported to the right neighbour $(i+1)$ and the rest is retained at site $i.$    
In other words,  from the site $i, $  $(1-\lambda)r_i$  fraction  of the mass 
is transported  to $(i+1).$ Here $r_i$ is a random number uniformly 
distributed in the interval $(0,1)$.

{\clr 
This asymmetric  sticky chipping model   can be  mapped  
to  the  generic one dimensional  mass  transport models discussed  
in Ref. \cite{FSS},    by considering  that  the  
amount of  mass $\mu$ that  is  transported 
from a site  to its right  neighbour is stochastic and  distributed as  
\beq
\phi(\mu|x) =  \left\{  \begin{array}{l l}
                  \frac{1}{(1-\lambda)x} &  0<\mu < (1-\lambda)x {\rm  }\cr
                  0 &  {\rm otherwise }
                \end{array}\right.\label{eq:phi}
\eeq 
where $x$ is the  initial mass of the departure site.  It has been shown \cite{FSS}  that 
the steady state  mass distribution $P_s(\{x_i\})$ 
of the generic model  is  factorized, $ i.e. ~P_s(\{x_i\})= \prod_i f(x_i),$  
only if  $\phi(\mu|x)$   has the  following form,  
\beq
\phi(\mu|x) =  \frac{ v(\mu) w(x-\mu)}{f(x)},\label{eq:phivw}
\eeq
where  $v$ and $w$ are arbitrary functions and 
\beq 
 f(x) = \left\{  \begin{array}{l l}
                 \int_0^x d\mu ~ v(\mu) w(x-\mu)& {\rm parallel ~~ update}\cr
                  w(x) &  {\rm random ~~sequential ~~ update. }
                \end{array}
\right. 
\label{eq:fssP(x)}
\eeq  

It is evident from Eqs. (\ref{eq:phivw}) and (\ref{eq:fssP(x)}) that  models with parallel
update can have a factorized steady state  only if  the dynamics  remain invariant 
when $(x-\mu)$ is transported instead of $\mu.$   For ASCM, however,  $(x-\mu)$  is 
distributed uniformly in the range $(\lambda x, x )$ (which is different from 
the range of $\mu$ in Eq. (\ref{eq:phi})) and thus the model can not 
have a factorized steady state  for parallel update unless $\lambda=0.$
For random sequential update, Eqs. (\ref{eq:phi})  and (\ref{eq:fssP(x)})  suggest that 
$f(x) = w(x) \sim x$; but  this can not be made consistent with Eq. (\ref{eq:phivw}) for any 
choice of $v.$ Thus random sequential update  can not give a factorized steady state 
for any $\lambda$ (including $\lambda=0$).  

In next two subsections we  
show that the  mass distribution $P(x),$ calculated perturbatively, 
matches  reasonably well with  the same obtained from Monte Carlo simulations.    
} 

\subsection{{\bf Parallel Update :}}
\label{AppI}~
\indent First let us consider the model with parallel update, where  all the 
sites are updated synchronously. The 
dynamics  can be written as 
\begin{equation}
x_i(t+1)= \lambda x_i(t)  +  (1-\lambda)(1-r_i) x_i(t) + 
(1-\lambda)r_{i-1}x_{i-1}(t)
\label{x'}
\end{equation}
for all $i=1,2,\dots ,L,$ where the first term on r.h.s. represents  the  mass that sticks
to the site $i,$ the second term 
corresponds to the  mass that is retained at the site $i$ after 
$(1-\lambda)r_{i}x_{i}(t)$  is transported to  $(i+1)$. 
The third term results from the mass that $i^{th}$ site  receives from $(i-1)$.

This model has a factorized steady state for $\lambda=0$ as the
distribution of transferred amount $\phi(\mu|x)= 1/x$  (from Eq. (\ref{eq:phi})) 
can be  written {\clr in the form} suggested by  Eq. (\ref{eq:phivw}) with $v(x)=1=w(x).$  
Correspondingly $f(x)= x,$ which results in $P(x) =  4x  e^{-2 x}.$  
This  special case of the model has been studied earlier  by Rajesh 
$et. al.$ \cite{Rajesh}. For any non-zero $\lambda,$ however, the steady state  
is not factorized.

\indent In order to find the steady state mass distribution $P(x)$ for generic $\lambda$ 
first we calculate the  moments. In the steady state, distribution of $x_i(t+1)$ 
is same as  the distribution of $x_i(t)$. Thus, one may  obtain 
$A^{(n)}\equiv\frac{\langle x^n\rangle}{(n+1)!}  $  from  Eq. (\ref{x'}) as
\begin{eqnarray}
A^{(n)}=\left[  n \blam -\lambda
+\lambda^{n+1}-\blam^{n+1}\right]^{-1} \sum_{k=1}^{n-1}(1-\lambda^{k+1}) \blam^{n-k}
A^{(k)} A^{(n-k)}
~~~~ \forall~  n\ge 2,\nonumber\\
\label{xn}
\end{eqnarray}
{\clr
where  we have used  
$ \langle  x_i^m x_{i-1}^n \rangle = \langle x_i^m \rangle   
\langle x_{i-1}^n \rangle,$ a mean field  approximation    that neglects 
all two point spatial correlations, and $\blam=(1-\lambda).$
Since $A^{(n)}$ depends on {\it all}  other $A^{(k)}$ with 
$k=1,2,\dots ,(n-1),$   it is usually difficult to get a general 
expression for $A^{(n)}$ from this recursion relation.  However, 
it is evident from  Eq. (\ref{xn})  that  if $y(n)$  is a solution for
$A^{(n)},$ $y(n)z^n$ is also a solution for any arbitrary  
$z.$  The arbitrary constant  $z$  must be chosen such that  the 
average mass of the system   has the  desired 
value $\langle x\rangle =1$ or in other words  
$A^{(1)}= \frac{1}{2}.$

First let  us  take $\lambda=0.$ In this case Eq. (\ref{xn}) reduces to
a very simple form
\beq  
 A^{(n)} =\frac {1}{n-1}\sum_{k=1}^{n-1} A^{(k)}  A^{(n-k)}   ~~~~~~ \forall ~ n\ge 2, \label{Eq.AP0s}
\eeq 
which  can be solved trivially by taking $A^{(n)}=1;$ thus a general solution is 
$A^{(n)}=z^n.$ The boundary condition $A^{(1)}=\frac{1}{2}$ now fixes $z=\frac{1}{2}$
and   thus $\langle x^n\rangle=\frac{(n+1)!}{2^n}.$ 
Corresponding 
steady state distribution function is then  
\begin{equation}
P(x)= \mathscr{L}^{-1} \left[
\sum_{n=0}^{\infty} \frac{(-s)^n}{n!}  \langle x^n\rangle  \right]= 4 x e^{-2x}. 
\label{p0s}
\end{equation}

For generic $\lambda,$ $\langle x^n\rangle$  can be calculated 
recursively   using Eq. (\ref{xn})   
starting from  $\langle x\rangle =1.$ The first few of them are, 
\begin{eqnarray}
\langle x^2\rangle=\frac{3 (\lambda +1)}{4 \lambda +2}\cr
\langle x^3\rangle=\frac{3 \left(\lambda ^2+3 \lambda +2\right)}{2 (2 \lambda
+1)^2}\cr
\langle x^4\rangle=\frac{15 (\lambda +1)^2 \left(\lambda ^2-\lambda +3\right)}{2
(2 \lambda
   +1)^2 \left(2 \lambda ^3-\lambda ^2+6 \lambda +3\right)}\cr
\vdots.
\end{eqnarray}
The moments $\langle  x^n \rangle$  as a function of $\lambda$ become messy 
with increasing $n$; obtaining  a general expression for $\langle  x^n \rangle$, 
and hence the  distribution $P(x)$, becomes 
practically impossible. It would be useful to obtain 
$P(x)$ perturbatively  which gives {\it all} the moments 
correctly {\clr (within  this mean field approximation)} up to  
some $n^{th}$  order  in $\lambda$.

To  proceed with the perturbation approach, we  first express 
$\langle x^n\rangle$ in a power series in $\lambda$ as  
in Eq. (\ref{Xs}) and  then 
equate the coefficients of different powers of $\lambda$  which gives  
a set of recursion  relations  for  $A^{(n)}_k.$  The recursion 
relation for  any $n^{th}$ order  perturbation in $\lambda$ can be 
solved using the boundary condition (\ref{Eq.bcA}). 
Once $A^{(n)}_k$ are known, we calculate $P_k(x)$ using Eq. (\ref{eq:Pks}).

In the $0^{th}$ order,  Eqs. (\ref{xn}) and (\ref{Xs})  result in       
\beq  
A_0^{(n)}= \frac {1}{n-1} \sum_{k=1}^{n-1}A_0^{(k)} A_0^{(n-k)}  ~~~~~~ \forall ~ n\ge 2, 
\label{A}
\eeq
which is indeed  same as Eq. (\ref{Eq.AP0s}). Thus to this order, as expected, we  get 
\begin{equation}
A^{(n)}_0= \frac{1}{2^n}, ~~{\rm and} ~~~ P_0(x)=4 x e^{-2x}. 
\label{p0}
\end{equation}

Next we proceed to calculate the $1^{st}$ order correction to $P(x).$ 
Comparing the  coefficients of $\lambda$ in  Eq.
(\ref{xn}) we   have 
\[A_1^{(n)}=\frac{1}{n-1}\sum_{k=1}^{n-1}[2A_0^{(k)} A_1^{(n-k)}-(n-k)A_0^{(k)}
A_0^{(n-k)}]~~~~~~ \forall ~ n\ge 2\]
which is needed to be solved using the boundary condition
$A_1^{(1)}=0$, from Eq. (\ref{Eq.bcA}).  This results in 
\begin{equation}
A_1^{(n)}=\frac{n}{2^n}\Big( 1-\gamma-\frac{\Gamma'(n+1)}{\Gamma(n+1) }\Big)~~~~~~ \forall ~ n\ge 0
\label{Eq.A1}
\end{equation}
where $\Gamma(x)$ stands for usual  gamma function and 
$\gamma$ is the {\it Euler constant}, 
\beq\gamma=0.57721\dots~~~.\label{gamma}\eeq  

Finally, using Eq. (\ref{eq:Pks}) we get
\begin{eqnarray}
P_1(x) =8e^{-2x}\left[x^2 \Big(1-g(x)\Big)+ x g(x)-\frac{1}{4}\right]
\label{p1}
\end{eqnarray} 
\beq \mbox{where}~ g(x)=\ln(2x)+\gamma \label{g}.\eeq 

\begin{figure}[h]
\centering
\includegraphics[width=15 cm]{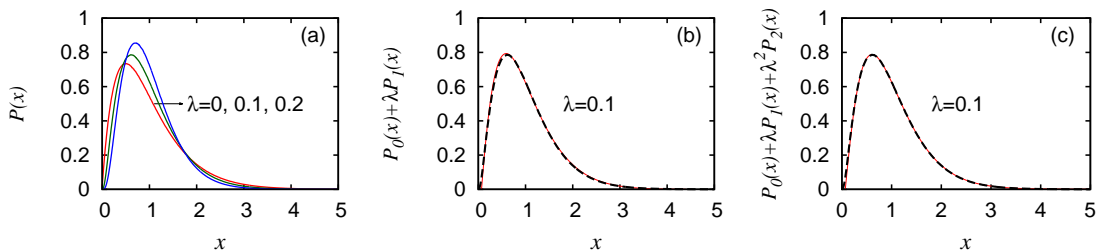}
\caption{ Steady state mass distribution $P(x)$ of ASCM with parallel update :
(a) $P(x)$ obtained from Monte Carlo simulations of the system of 
size $L=1000$,  for $\lambda=0, 0.1,$ and $0.2$.
(b) and (c) compare $P(x)$  for $\lambda=0.1$ with the $1^{st}$  and 
$2^{nd}$ order  perturbation results (dashed line) respectively.
}
\label{f1}
\end{figure}
~\\

One can proceed  in a similar way to calculate  higher order corrections 
to $P(x).$ The corrections up to  $2^{nd}$ order  in $\lambda$  
are listed in the Appendix.

To check that the  perturbation  results  for $P(x)$  agree well with 
the actual  steady state mass distribution  we simulate the model on a 
one dimensional lattice of size $L=1000$    for $\lambda=0,0.1,$ and $0.2$  
(shown in  Fig. \ref{f1}(a)).  The distribution for $\lambda=0.1$ (solid line) 
is compared    with the  perturbation results (dashed line) 
up to $1^{st}$ and $2^{nd}$ order  respectively in 
Figs. \ref{f1}(b)  and (c).  An excellent agreement of  $P(x)$ with 
the simulation  results indicate that   the  two  point correlations among 
sites are  indeed very small.   

\subsection{{\bf Random sequential update :}}~
\indent In this subsection we study the \textit{asymmetric sticky chipping
model} (ASCM) with random sequential update where 
$(1-\lambda)$ fraction of the mass $x_i$  from a randomly chosen site $i$ is 
chipped off; a part of it, $(1-\lambda)r_i x_i$  is then transported to the 
right neighbour $(i+1)$ and  the  rest  is returned  to the departure site $i$. 
As usual, $r_i$ is distributed uniformly in the interval $(0,1)$.
{\clrr The time increment  associated with each update  is $\Delta t=1/L$; 
in other words a unit Monte Carlo sweep (MCS)  corresponds to the 
update  of $L$ sites. }

Like the  parallel update, here too one can obtain all the moments,  

\begin{equation}
\langle x_i^n\rangle=\frac{1}{2}\left\langle \Big(\lambda+(1-\lambda)(1-r_i)\Big)^n
x_i^n\right\rangle+\frac{1}{2}
\left\langle\Big(x_i+(1-\lambda)r_{i-1}x_{i-1}\Big)^n\right\rangle,
\label{xnar}
\end{equation}
for all $i=1,2,\ldots,L.$ {\clr 
{\clrr The factors $\frac 1 2$  here represent the fact that the probability that
a randomly chosen site acts as a departure site  is $\frac 1 2$ and the probability
of acting as a receiving site is the same.}
Now, using  a  mean 
field approximation $\langle  x_i^m x_{i-1}^n \rangle = \langle x_i^m \rangle   
\langle x_{i-1}^n \rangle$, $A^{(n)} \equiv \frac{\langle x^n 
\rangle}{(n+1)!}$ can be written  in terms of  $A^{(k)}$ as
\begin{eqnarray}
A^{(n)}=\left[ 
 n \blam -\lambda
+\lambda^{n+1}-\blam^{n+1} \right]^{-1} \sum_{k=1}^{n-1}(n-k+1) \blam^{k+1}A^{(k)} A^{(n-k)}
~~~~ \forall~  n\ge 2,\nonumber\\
\label{Eq.xnASCMR}
\end{eqnarray}
where $\blam= (1-\lambda).$ For $\lambda=0$ this recursion relation 
reduces to  a simpler form, 
\beq  A^{(n)}=\frac{1}{n-1}\sum_{k=1}^{n-1}(n-k+1)A^{(k)} A^{(n-k)}
 ~~~~ \forall ~ n\ge 2 \label{Eq.AnAR},\eeq
which can be solved using the boundary condition  $A^{(1)} = \frac 1 2,$
\beq
A^{(n)} =  \frac{(2n)!}{2^n n! (n+1)!}  ~~~~\forall ~n\ge0.
\eeq
In other words we have $\langle x^n \rangle = \frac{(2n)!}{2^n n!}.$  Thus the 
steady state distribution for $\lambda=0$ is  
\beq P(x)=\mathscr{L}^{-1} \Big[  \sum_0^\infty  \frac{ (-s)^n}{n!} 
\langle x^n \rangle \Big]= \frac{1}{\sqrt{2
\pi x}}e^{-x/2}.\eeq
Note that $P(x)$ is same as the mass distribution of the 
$\lambda=0$ case, obtained earlier \cite{Rajesh} for random sequential update.

We must mention that, a closed form expression of the steady state distribution 
can also be obtained for a special value of  stickiness $\lambda=\frac{1}{2}.$  
In this case Eq. (\ref{Eq.xnASCMR}) gives
\begin{equation}
A^{(n)}=\frac{1}{n-1}\sum_{k=1}^{n-1}\frac{n-k+1}{2^k}A^{(k)} A^{(n-k)}~~~~\forall~ n\ge2.
\end{equation}
This equation has the trivial solution $A^{(n)}=1,$   similar to the case 
when the system evolves following  a parallel update with  $\lambda=0$ 
(discussed in the previous subsection).  Thus, in the steady state  we have 
the same distribution as obtained in Eq. (\ref{p0s}), 
\beq P(x)=4x e^{-2x}. \label{Eq.phf}\eeq
In  Fig. \ref{f2}(a) we have compared this result  with the steady state distribution 
obtained from the Monte Carlo simulation of the system  with $\lambda=\frac{1}{2}.$

 For  any arbitrary $\lambda$, however,  obtaining the solution of Eq. (\ref{Eq.xnASCMR}) 
is not easy and we resort to the perturbation approach. As usual, first we 
expand $A^{(n)}$ as  a power series in $\lambda$
as  done in Eq. (\ref{Xs}) and then equate the coefficients of different powers of 
$\lambda$ in Eq. (\ref{Eq.xnASCMR}).  To $0^{th}$ order, $i.e.$  by equating  the 
 terms independent of $\lambda$, we have 
\beq A_0^{(n)}=\frac{1}{n-1}\sum_{k=1}^{n-1}(n-k+1)A_0^{(k)} A_0^{(n-k)} 
~~~~~~ \forall ~ n\ge 2.\eeq
As expected, this equation is same as Eq. (\ref{Eq.AnAR}) and 
correspondingly, 
\beq
A_0^{(n)} =  \frac{(2n)!}{2^n n! (n+1)!} ~~ {\rm and} ~~P_0(x)=\frac{1}{\sqrt{2\pi x}}e^{-x/2}.
\label{eq:A0ASCMrs}
\eeq

Now let us proceed to the first order perturbation calculations. Collecting the coefficients 
of $\lambda$ in Eq. (\ref{Eq.xnASCMR}) one gets
\begin{eqnarray}
A_1^{(n)}&=&\frac{1}{n-1}\sum_{k=1}^{n-1}\Big[-(n-k+1)(k+1)A_0^{(k)}
A_0^{(n-k)}+(n-k+1)\nonumber\\
&&(A_0^{(k)} A_1^{(n-k)}+A_1^{(k)} A_0^{(n-k)})\Big] ~~~~~~ \forall ~ n\ge 2.
\label{eq:A1ASCMrs}
\end{eqnarray}
This recursion relation  can be  solved using 
the generating function, 
\beq 
V_k(s)=\sum_{n=0}^\infty (-s)^n A_k^{(n)}.
\eeq
In terms of $V_0(s)$ and $V_1(s)$, Eq. (\ref{eq:A1ASCMrs}) can be 
written as a differential equation  
\begin{eqnarray*}
&&-s V_0(s)V_1'(s)+2s V_1'(s)-s V_0'(s)V_1(s)-2V_0(s)V_1(s)+V_1(s)\\&&
+sV_0(s)V_0'(s)
+s^2V_0'^2(s)-V_0(s)+1=0.
\end{eqnarray*}
Since  $V_0(s)=(\sqrt{2 s+1}-1)/s$ is known from Eq. (\ref{eq:A0ASCMrs}) 
we can solve the above equation for $V_1(s)$  using 
the boundary condition $V_1'(0)=0$ (obtained from Eq. (\ref{Eq.bcA})), 
\beq V_1(s)=-\frac{2 s \left(\tilde{s}-3\right)+4
\left(\tilde{s}-1\right)+2\left(s-\tilde{s}+1\right) \ln (\tilde{s})}{2
   s \tilde{s}}.\label{Eq.ASCMRV1}\eeq
Here $\tilde{s}=\sqrt{1+2s}$.
Thus following Eq. (\ref{eq:Pks}), we get
\begin{eqnarray}
P_1(x)&=&\mathscr L^{-1}\left[V_1(s)+s V_1'(s)\right]\nonumber\\
& =&\frac{1}{4\sqrt{2 \pi x}} e^{-x/2} \Big[
 -x \Big( g(x)+2 \Big) + g(x)+4 \Big] + \frac 1 2 e^{-x/2} -2
\delta(x)
\label{Eq.ASCMRP1}
\end{eqnarray}
where $g(x)=\ln(2x)+\gamma$.
In the above expression the term $-2\delta(x)$ is needed to ensure the condition
$\int_0^\infty dx~P_1(x)=0$.\\

\begin{figure}[!h]
\centering
\includegraphics[width=15 cm]{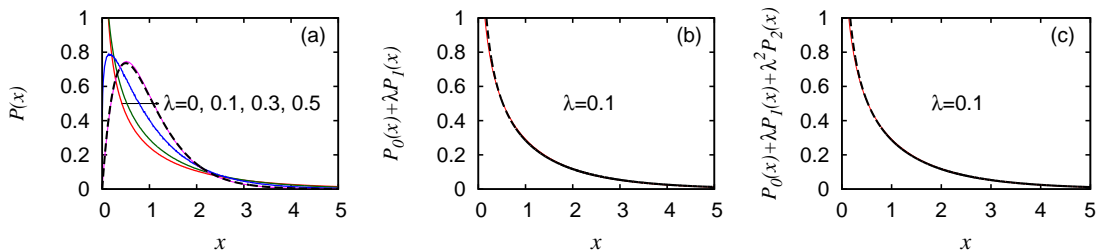}
\caption{The steady state distribution $P(x)$ of ASCM with random sequential 
update:  (a) obtained from Monte Carlo 
simulations of system  size $L=1000$  for $\lambda=0, 0.1, 0.3,$ and $0.5$. 
The distribution for  $\lambda=0.5$ is  compared  with  the 
analytical result Eq. (\ref{Eq.phf})  (shown as dashed line).
In  (b) and (c)   we have  compared the distribution  for $\lambda=0.1$ 
with the $1^{st}$  and  $2^{nd}$ order perturbation results (dashed line) respectively.}
 \label{f2}
\end{figure}
~\\

There is no particular difficulty in proceeding for higher order 
perturbations in $\lambda,$ except that the expressions are lengthy. 
We have listed the $P_k(x)$ for $k=0,1,2$ in the  Appendix.

In Fig. \ref{f2}  we have  compared  these  perturbation 
results with  the  Monte Carlo simulation of the  dynamics on a 
one dimensional periodic system of size $L=1000.$   
The steady state  distribution for $\lambda=0, 0.1, 0.3,$ and $0.5$ 
are shown in Fig. \ref{f2}(a). In the same figure we compare the distribution function
for $\lambda=0.5$ with Eq. (\ref{Eq.phf}) (dashed line). The distribution for $\lambda=0.1$ (solid line) 
is compared  with the  perturbation results (dashed line) 
up to $1^{st}$ and $2^{nd}$ order  respectively in 
Figs. \ref{f2}(b)  and (c).

\section{Symmetric Sticky Chipping Model (SSCM) :}
\label{SSCM}~
\indent In this section we study the symmetric version of the model, namely
SSCM, where  $(1-\lambda)$ fraction of the mass $x_i$ at site $i$ is 
chipped off and distributed randomly among  the 
neighbours $(i\pm1)$; the right neighbour receives $\mu_R=r_i(1-\lambda)x_i$
and the left one receives the rest $\mu_L= (1-r_i)(1-\lambda)x_i$ . 
Here again  $r_i$ is a random number uniformly distributed in the interval $(0,1)$. 

 The   criterion (\ref{eq:phivw}) for  asymmetric mass transport models to 
have  factorized steady state \cite{FSS}, does not straight forwardly 
extend to the symmetric case. {\clrr However, to have  factorized steady state 
for a chipping model on an arbitrary  graph, it is  {\it sufficient} that its  
chipping kernel at each site  has a product form \cite{FSS2}
similar to  Eq. (\ref{eq:phivw}). In SSCM, this condition  translates to 
\begin{equation}
\phi_{sym}(\mu_L, \mu_R|x) = \frac{u(\mu_L) v(\mu_R) w(x-\mu_L-\mu_R)}
{ \int d\mu_L d \mu_R  u(\mu_L) v(\mu_R) w(x-\mu_L-\mu_R)}. \label{eq:phi_sym}
\end{equation}
The chipping kernel  in this model  is 
$$ \phi_{sym}(\mu_L, \mu_R|x) =  \delta\left((1-\lambda)x-\mu_L-\mu_R\right) \phi(\mu_L|x),$$
where  $\phi(\mu_L|x)$ is the  same distribution   given  by 
Eq. (\ref{eq:phi}).  It is evident   $\phi_{sym}(\mu_L, \mu_R|x)$ can be  cast into the 
the form (\ref{eq:phi_sym})  when $\lambda=0,$ by taking  the functions $u=1=v$ (similar to 
ASCM) and $w= \delta(x-\mu_L-\mu_R);$   thus  the steady state is factorized. 
} For $\lambda\ne0,$   however, such product form  does not  exist and  we  proceed  
to calculate  the  steady state 
distribution  perturbatively  for  both parallel and random 
sequential updates.

\subsection{{\bf Parallel Update :}}~
\indent In this subsection  we study the model where all the sites are updated
parallely (synchronously) using the 
dynamics mentioned above. Explicitly,
\begin{equation}
x_i(t+1)=\lambda x_i(t)+(1-\lambda)[r_{i-1}x_{i-1}(t)+(1-r_{i+1})x_{i+1}(t)] 
\label{Eq.xi}
\end{equation}
where, the $1^{st}$ term on the r.h.s. represents the mass that sticks to the
site $i$ during the update, the 
$2^{nd}$ and the $3^{rd}$ term  there correspond to the mass which  $i^{th}$ site 
receives from
its neighbours $i\mp 1$  respectively.  Thus the steady state probability that  the 
site $i$ has mass $x$ is given by, 

\begin{eqnarray}
P(x)&=&\int_0^{\infty}dx_{i-1}\int_0^{\infty}dx_i\int_0^{\infty}dx_{i+1}\int_0^1 dr_{i-1} 
\int_0^1 dr_{i+1}  
 P(x_{i-1})P(x_i)P(x_{i+1})\cr
&&
\delta\Big(x-\lambda x_i-(1-\lambda)[r_{i-1}x_{i-1}+(1-r_{i+1})x_{i+1}]\Big),
\end{eqnarray}
where we have  used a mean field approximation 
$\langle x_{i-1}^l  x_i^m x_{i+1}^n \rangle =   
\langle x_{i-1}^l \rangle  \langle x_i^m \rangle
\langle x_{i+1}^n \rangle.$  In other words  
three  and lower  order  spatial  correlations  are  ignored.

Here we intend to follow approach II. First let us  express the above 
equation in terms of  $Q(s)$,  which is the 
Laplace transform  of $P(x)$; 
\begin{eqnarray}   
Q(s)=Q(\lambda s)V^2\Big((1-\lambda)s\Big),
 \label{Qsg}
\end{eqnarray}
where $V(s)$ is defined by
\beq
V(s)=\int_0^1 dr~ Q(s r), ~~~\mbox{or}~~~ Q(s)= V(s) + s V'(s).
\label{Eq.QV}
\eeq
In fact Eq. (\ref{Qsg})  can be written  in terms of 
the function $V$ as 
\begin{eqnarray}   
 V(s) + s V'(s) = \left[  V(\lambda s) + \lambda s V'(\lambda s)\right] V^2\Big((1-\lambda)s\Big).
\label{Vsg}
\end{eqnarray}

Let us begin with  $\lambda=0$. In  this case, Eq. (\ref{Vsg}) takes the form \beq s V'(s)-V(s)^2+V(s)=0, \label{Eq.v0s}\eeq
where we have used $V(0)=1,$ from Eq. (\ref{eq:avrx=1}).
The solution of this differential equation,  with the  
boundary condition $V'(0)=-1/2$  which corresponds 
to  a fixed average   mass $\langle x\rangle=1$ gives, 
$$V(s)=\frac{2}{s+2},  {\rm~
and~ thus~}  P(x)=\mathscr {L}^{-1}\left[ V(s) +s V'(s)\right]=4x e^{-2 x}.$$
This expression  is identical to  $P(x)$ obtained in case of ASCM  
with parallel update. In the following we argue that the dynamics of  
SSCM  for $\lambda=0$ is equivalent to   that of ASCM :
\begin{eqnarray*}
x_i(t+1)=r_{i-1}x_{i-1}(t)+(1-r_{i})x_{i}(t)  & ~~~~~~{\rm ASCM} \cr 
x_i(t+1)=r_{i-1}x_{i-1}(t)+(1-r_{i+1})x_{i+1}(t)  & ~~~~~~{\rm SSCM}.
\end{eqnarray*}
Since ASCM  has a product measure, the steady state  remains invariant if 
the dynamics is changed to a mean field dynamics 
$x_i \to (r_{j}x_{j} + (1-r_{k})x_{k}),$  where for each  $i,$  
$j$ (and $k$) are chosen randomly from  the set $\{1,2,\dots ,L\}$ without  replacement,
$i.e.$ all the sites 
receive  {\it exactly two fragments},  $r_j$ fraction of  $x_j$ and $(1-r_{k})$ fraction 
of $x_{k}$.  Clearly  the choice $j= (i-1)$ and $k=(i+1)$   corresponds to  SSCM. 
Thus, SSCM with $\lambda=0$  has a factorized steady state  with  mass distribution
$P(x)$ same as ASCM.

These arguments do not extend to $\lambda\ne 0$ case, as  the steady state 
of the  corresponding  asymmetric  model is not  factorized.  We proceed with 
the perturbation approach.  First  let us  expand  $V$ in the Taylor series about $\lambda=0$; for any arbitrary 
$a, b$ 
 \beq V\Big((a+b \lambda)s\Big)=\sum_{m=0}^\infty \frac{(b\lambda
s)^m}{m!} \frac{d^m V(s)}{ds^m}\Bigg|_{s\to as} .\label{VT}\eeq
\noindent Each of the $V(s)$ and their derivatives are then expanded, similar to Eq. (\ref{Qs}), 
as \footnote{Note that $V(s)$ is also a function of $\lambda$, which we have dropped 
for notational convenience.}
\beq V(s)=\sum_{k=0}^\infty \lambda^k V_k(s).\label{VsT}\eeq
Now using  Eqs. (\ref{VT}) and (\ref{VsT}) in Eq. (\ref{Vsg}) and collecting the
coefficients of different powers of 
$\lambda$, order by order, one can obtain a set of differential equations in
terms of $V_k(s),$ which can be solved  using the boundary conditions
\begin{eqnarray}
k=0~~~:V_0(0)=1,~~~V_0'(0)=-\frac{1}{2}\nonumber\\
k\ne 0~~~:V_k(0)=0,~~~V_k'(0)=0.
\label{bcV} 
\end{eqnarray}
These  boundary conditions on $V$ are  same as  Eqs. (\ref{C0}) and (\ref{Eq.pkb}),
which ensure that  $P(x)$ is normalized and the  average mass  of the system is unity.
Interestingly,  using  Eq. (\ref{Qk})  in Eq. (\ref{Eq.QV})  one gets 
\beq V_k(s)=\sum_{n=0}^\infty(-s)^n A_k^{(n)},\label{VkAk}\eeq
which implies that 
$V_k(s)$  is   simply the generating function of $A_k^{(n)}$ used 
in approach I. 

Following the perturbation approach,   we use Eqs. (\ref{VT}) and (\ref{VsT})  
in  Eq. (\ref{Vsg}) and equate the  terms  which are independent of $\lambda$
and get 
\[s V_0'(s)-V_0(s)^2+V_0(s)=0,\]
which is same as Eq. (\ref{Eq.v0s}) and we get
\[V_0(s)=\frac{2}{s+2},  ~~{\rm and}~~ P_0(x)=4x e^{-2 x} .\]

Next we move to the first order perturbation. Collecting the coefficients of $\lambda$ in Eq. (\ref{Vsg}),
\[s(s+2)^3V_1'(s)+(s+2)^2(s-2)V_1(s)+4s^2=0\]
which, along with  the  boundary condition $V_1'(0)=0$ (from Eq. (\ref{bcV})),
results in \[V_1(s)=-\frac{4 s \Big(\ln(s+2) -\ln(2)\Big)}{(s+2)^2}.\]
Thus we obtain
\begin{eqnarray*}
P_1(x)=\mathscr {L}^{-1}\left[ V_1(s) +sV_1'(s)\right]=16e^{-2x}\left[x^2\Big(1-g(x)\Big)+xg(x)-\frac{1}{4}\right]
\end{eqnarray*}
where $g(x)=\ln(2x)+\gamma$.

\begin{figure}[h!]
\centering
\includegraphics[width=15 cm]{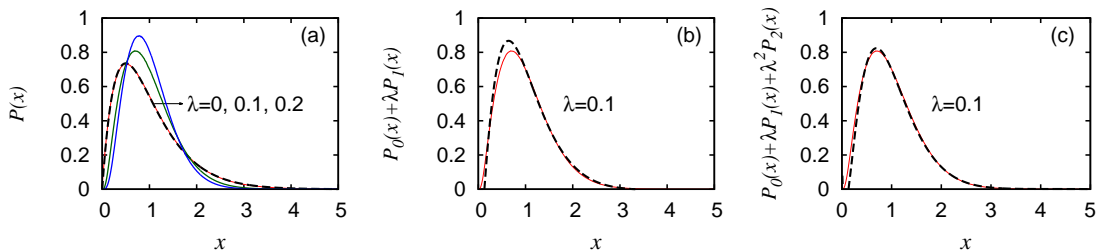}
\caption{ SSCM with parallel update : (a) $P(x)$ obtained from Monte Carlo simulations of  
the model on a one dimensional lattice  of 
size $L=1000$ for $\lambda=0,0.1,$ and $0.2$.  The distribution  for  
$\lambda=0$  is compared with the  exact result $P(x) = 4x e^{-2x}$ 
(dashed line). (b) and (c)  compare $P(x)$  obtained   from  the $1^{st}$  and  $2^{nd}$ order 
 perturbation (dashed line) respectively with the same obtained from  simulations, for  $\lambda=0.1$.}
\label{f3}
\end{figure}
Following the same procedure one can obtain other higher order corrections to 
$P(x).$  In the Appendix, we have listed these correction terms up to 
$2^{nd}$ order. As  described in  Fig. \ref{f3},  the  perturbation  results  
for $P(x)$  agree well with  the actual  steady state mass distribution  
obtained from  the Monte Carlo  simulation.   {\clr In Fig. \ref{f3}(a)   we have 
shown  the simulation results for $\lambda=0,0.1,$ and  $0.2.$} In the same 
figure we compare the result for $\lambda=0$ with the analytically obtained result 
(dashed line). In Figs. \ref{f3}(b) and (c) the distribution for $\lambda=0.1$ (solid line) 
is compared   with the  perturbation results (dashed line) 
up to $1^{st}$ and $2^{nd}$ order  respectively.  Clearly, as  expected, 
the distribution  matches better with the  
simulation results as we  go to 
higher order.

\subsection{{\bf Random sequential  update :}}~
\indent In this subsection we study the {\it symmetric sticky chipping model}
using random sequential update. 
At each step, a  site  is chosen randomly and  from this site $i,$ 
$(1-\lambda)$ fraction of the mass $x_i$ is chipped off.  From the
chipped off mass $(1-\lambda)x_i,$  $r_i$  fraction is transported to 
the right neighbour,  and  the rest goes to  the left.  
{\clrr With update of a single site, the time is increased by   $\Delta t= 1/L.$ }

The moments $\langle
x_i^n\rangle$ can be expressed as
\begin{eqnarray} 
\langle x_i^n\rangle&=&\frac{1}{3}\Big\langle (\lambda
x_i)^n\Big\rangle+\frac{1}{3}\left\langle\Big(x_i+r_{i-1}(1-\lambda)
x_{i-1}\Big)^n\right\rangle \cr
&& ~~~~~~~~~+\frac{1}{3}\left\langle \Big( x_i+(1-r_{i+1})(1-\lambda) x_{i+1}\Big) ^n\right \rangle \nonumber\\
&=&\frac{1}{3}\Big\langle (\lambda
x_i)^n\Big\rangle+\frac{2}{3}\left\langle\Big(x_i+r_{i-1} (1-\lambda)x_{i-1}\Big)^n\right\rangle.
\label{Eq.SSCMRxn}
\end{eqnarray}
The  factors $\frac  1 3$   come from the fact that 
{\clrr during each update, one site transports and two other sites  
receive the mass. Thus, probability  that a site 
acts as a departure site is $\frac  1 3$  and the probability 
that it acts as a receiving site is $\frac 2 3.$ }
%

Now, consider the case $\lambda=0$. Then, Eq. (\ref{Eq.SSCMRxn}) takes the form
\[\langle x_i^n\rangle=\frac{1}{3}\delta_{n,0}+ \frac{2}{3}\langle(x_i+r_{i-1} x_{i-1})^n\rangle.\]
{\clrr The additional term $\frac{1}{3}\delta_{n,0}$ appears from  the fact that  for $n=0,$ 
the first  term on the r.h.s. of Eq. (\ref{Eq.SSCMRxn}) 
gives  $\frac 1 3$ for any arbitrary  $\lambda$ except  $\lambda=0.$}

Assuming that  all the two point  correlations are zero in the steady state, 
$i.e.$ $\langle  x_i^m x_{i-1}^n \rangle = \langle x_i^m \rangle   
\langle x_{i-1}^n \rangle,$  one can write  $ A^{(n)}\equiv\frac{\langle x^n\rangle}{(n+1)!}$ as
\begin{equation}
A^{(n)}=\frac{2}{n-1}\sum_{k=1}^{n-1}(n-k+1)A^{(k)}A^{(n-k)}  ~~~~~\forall~ n\ge2.
\label{Eq.SSCMRA0}
\end{equation}
This equation  can be converted to a differential equation  
by using $V(s)=\sum_{n=0}^\infty (-s)^n
A^{(n)},$ the generating function of $A^{(n)},$  
\[V'(s)=\frac{1-3V(s)+2V^2(s)}{s\Big(3-2V(s)\Big)}.\]
For the  usual boundary condition $V(0)=1,~V'(0)=-1/2$, 
the solution to  the above equation is   
\[V(s)=\frac{2 s-1+\sqrt{4 s+1}}{4 s}\]
which results in
\begin{equation}
P(x)=\mathscr{L}^{-1}\left[V(s)+s
V'(s)\right]=\frac{1}{4\sqrt{\pi x}} e^{-x/4}+\delta (x).
\label{Eq.SSCMRP0}
\end{equation}
In the above expression the term $\delta(x)$ is needed to assure the
normalization of $P(x)$. 
\begin{figure}[h]
\centering
 \includegraphics[width=10cm]{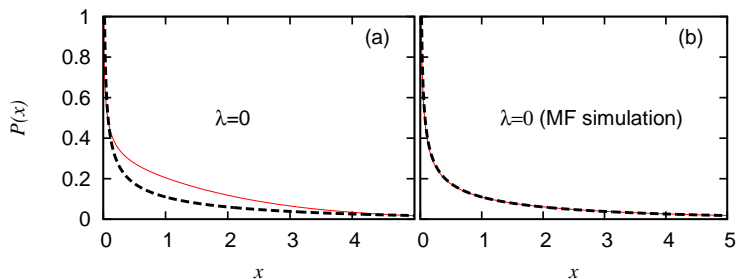}
\ \caption{Comparison of Eq. (\ref{Eq.SSCMRP0}) (dashed line) with $P(x)$ obtained
from  Monte Carlo simulations (solid line) of SSCM with random sequential
update  for  $\lambda=0$  and $L=1000$.  (a) The actual model. (b) The model with mean field 
(MF) dynamics.
} 
\label{f.SSCMR}
\end{figure}

In Fig. \ref{f.SSCMR}(a) we have compared this 
result  with the  Monte Carlo simulation of the model with $\lambda=0$.  
{\clrr The right panel (b)   shows $P(x)$ obtained from  Eq. (\ref{Eq.SSCMRP0})
along with the  simulation results of the  model with a mean field (MF) dynamics, 
{\it i.e.},  $r_i$ and $(1-r_i)$ fraction of the  chipped off mass 
$(1-\lambda) x_i$ are transported from site $i$ to two arbitrary sites  
instead of being transported to the neighbours. 
Clearly Eq. (\ref{Eq.SSCMRP0}) is consistent  with  simulation results of 
the MF dynamics emphasizing  that  Eq. (\ref{Eq.SSCMRP0}) correctly 
describes the   mean field distribution. On the other hand  
the same mass distribution  obtained from Monte Carlo simulation of  
the actual model  deviates substantially (Fig. \ref{f.SSCMR}(a)). }
This discrepancy, which originates from  the mean field 
approximation used here (that ignores  all the two point correlations),
can not be healed  by adding  perturbative  correction  terms.
We do not proceed further in this case.

\section{Summary and conclusion :}
\label{summary}~
\indent In this article we have studied  the conserved mass transport
process in presence of stickiness, characterized by a parameter $\lambda$.  
The model in  one dimension  evolves using a parallel or random sequential 
update rule, where  a fixed fraction ($ i.e. ~~ (1-\lambda)),$ of the mass 
from a  site is chipped off and then distributed randomly 
among the site and its neighbours. In the asymmetric version ASCM, the chipped 
off mass is distributed among the site and  its right neighbour, whereas 
in SSCM  it is distributed among both the  neighbours.

For nonzero $\lambda$ the  steady state  distribution  of these 
models   do not have factorized form. We introduce a perturbation approach 
to obtain an  approximate mass distribution function  $P(x)=\sum_{k=0}^
\infty \lambda^k P_k(x)$,  and  provide  explicit form  of 
$P_k(x)$ up to $2^{nd}$ order in $\lambda.$  In  all cases except   SSCM with 
random sequential  update,   the perturbation results agree quite well with the 
distribution obtained from  the Monte Carlo simulation of the model, even though 
we have used a mean field approximation which  ignores two or three  point spatial 
correlations.

Interestingly, the steady state distribution for the following  
three cases, 
(a) ASCM with  parallel update  and $\lambda=0,$ 
(b) ASCM with  random sequential update  and $\lambda=\frac 1 2 ,$ and 
(c) SSCM with  parallel update  and $\lambda=0,$ 
are identical, $P(x) = 4x e^{-2x}.$   It turns out  that  product measure 
is exact  only for models (a) and (c).  
 
 In absence of any general formalism,  calculating the  exact steady state distribution
for  non-equilibrium models  is  not always possible. 
The perturbation approach we discussed here is quite general and can be used  
in models  with  some {\it small} parameter to  obtain   
steady state distribution  analytically  within a mean field approximation
that  ignores only two point correlations   in all cases, except for 
SSCM with parallel update where both two and three point correlations are ignored.  
 \\~

{\it Acknowledgements:}   We would like  to thank the referees  for their  useful 
comments and suggestions, and U. Basu for careful reading of the manuscript.

\newpage
 \section*{Appendix :}~
The  steady state distribution of the models studied  here with
parallel (p) and random sequential (rs) updates,
using  the perturbation approach,  are summarized below.\\
\begin{tabular}{|l|l|}
\hline
Models& $P(x) = P_0(x) + \lambda P_1(x) + \lambda^2 P_2(x) + \dots$\\
\hline
ASCM (p)   & $P_0(x)=4 x e^{-2x}$  \\  
& $P_1(x) =8e^{-2x}\left[x^2 \Big(1-g(x)\Big)+ x g(x)-\frac{1}{4}\right]$\\
& $P_2(x)=8 e^{-2 x}
\Big[x^3\Big(g^2(x)-2g(x)+\frac{18-\pi^2}{6}\Big)-x^2\Big(\frac 5 2 g^2(x)-2g(x)$\\
& $~~~~~~~+ \frac{48- 5 \pi^2}{12}\Big)
+x\Big(g^2(x)+ \frac 3 2 g(x)-\frac{\pi^2}{6}\Big)-  \frac 1 4 \Big( g(x)+ 1\Big)  \Big]+2\delta
(x)$   \\
 
\hline
ASCM (rs) & $P_0(x)=\frac{1}{\sqrt{2
\pi x }}  e^{-x/2}$   \\  
 & $P_1(x) =\frac{1}{4\sqrt{2 \pi x}} e^{-x/2} \Big[
 -x \Big( g(x)+2 \Big) + g(x)+4 \Big] + \frac 1 2 e^{-x/2} -2
\delta(x)$\\
 & $P_2(x)=\frac{1}{32\sqrt{ 2 \pi x^3} }e^{-x/2}\Big[
x^3  \Big( g^2(x)  +  4 g(x) + \frac{20-\pi^2}{2} \Big) 
-4x^2 \Big( g^2(x)+ 6 g(x) $\\
&$~~~~~~~~~~~~~~~~~~~~~~~~~~~$ 
$ + \frac{43-3\pi^2}{6}\Big)+x \Big( g^2(x)+ 12 g(x)  -\frac{20+3 \pi^2}{6}\Big) -10 \Big]$\\
&$~~~~~~~~~~$ $-\frac{1}{8}e^{-x/2}
\Big[ x  \Big(  g(\frac x 4)+3\Big) -2g(\frac x 4)-10 \Big]$
$+\frac2 3  \Big[ \delta '(x) -  2\delta (x)\Big] $   \\
 
\hline
SSCM  (p) & $P_0(x)=4 x e^{-2x}$  \\   
& $P_1(x) =16e^{-2x}\left[x^2\Big( 1- g(x)\Big) + x g(x)-\frac{1}{4}\right]$\\
   & $P_2(x)=32 e^{-2 x} \Big[ x^3 \Big( g^2(x)-2g(x)+
\frac{33- 2\pi^2 }{12} \Big) - x^2\Big( 3 g^2(x)-2 g(x) $\\
&$~~~~~~~$  $ -H(x)+\frac{21- 2\pi^2}{6}\Big)+ x\Big( \frac 3 2 g^2(x)+g(x)-H(x)-\frac{\pi^2}{12}\Big)
   $\\
&$~~~~~~~$ $ - \frac 1 4 \Big( g(x)+ 1 \Big) \Big]- 8 \Gamma(0,2x)( 2x-1)   +6 \delta (x)$ \\ 

\hline
SSCM (rs)   & $P_0(x)=\frac{1}{4\sqrt{\pi x}} e^{-x/4}+\delta (x)$  \\  
\hline
\end{tabular}

Here,  $g(x)= \ln(2x) +\gamma,$ and $H(x) = \sum_{m=1}^\infty [\frac{2}{m}+ 
\gamma+\frac{\Gamma'(m)}{\Gamma(m)} - g(x) ]\frac{(2x)^m}{mm!}.$ \\$\Gamma(0,x)$ is  
the incomplete  gamma function and  Euler constant $\gamma=0.57721...$

\end{document}